\documentclass[lettersize,journal]{IEEEtran}
\usepackage{amsmath,amsfonts}
\usepackage{algorithmic}
\usepackage{array}
\usepackage[caption=false,font=normalsize,labelfont=sf,textfont=sf]{subfig}
\usepackage{textcomp}
\usepackage{stfloats}
\usepackage{url}
\usepackage{verbatim}
\usepackage{graphicx}
\def\BibTeX{{\rm B\kern-.05em{\sc i\kern-.025em b}\kern-.08em
    T\kern-.1667em\lower.7ex\hbox{E}\kern-.125emX}}
\usepackage{balance}
\begin{document}

\title{The 2023 Development of Room-Temperature Ambient-Pressure Superconductor: Vision and Future Trend of Power Systems}

\author{Yi Yang,~\IEEEmembership{Graduate Student Member,~IEEE,} Chenxi Zhang,~\IEEEmembership{Graduate Student Member,~IEEE,} Xinlei Wang,~\IEEEmembership{Graduate Student Member,~IEEE,}, Jing Qiu,~\IEEEmembership{Senior Member,~IEEE,} Jinjin Gu~\IEEEmembership{Graduate Student Member,~IEEE,} and Junhua Zhao,~\IEEEmembership{Senior Member,~IEEE,}
\thanks{Manuscript created October 2020; This work was developed by the IEEE Publication Technology Department. }}

\markboth{The 2023 Development of Room-Temperature Ambient-Pressure Superconductor: Vision and Future Trend of Power Systems}%
{}

\maketitle

\begin{abstract}
Room-Temperature Ambient-Pressure Superconductor (RTAPS) can achieve superconducting properties at room temperature and normal atmospheric pressure, eliminating the power system's transmission loss and enhancing power systems efficiency. This paper investigates the comprehensive implications and prospective applications of the recently discovered RTAPS, LK-99, in modern power systems. It explores the potential of RTAPS in reshaping modern power systems paradigms, providing the vision of future RTAPS-based power systems. Although debate surrounds its industrial implementations, RTAPS's benefits for electricity transmission, grid flexibility, improved energy storage, and renewable energy integration could be unprecedented. The paper delves into underlying opportunities and challenges, including RTAPS-based power flow methods evolution, security redefinition, computational efficiency, cost implications, and innovative market transaction forms to facilitate renewable energy competitiveness.
%In traditional power system planning and operation problems, the system optimization time typically depends on the complexity of power flow calculations, and line impedance can cause a series of issues, such as transmission power loss and voltage drop. Room-Temperature Ambient-Pressure Superconductor (RTAPS) is a material that can achieve superconducting properties at room temperature and normal atmospheric pressure. The discovery and commercialization of RTAPS could instigate significant changes in existing power system planning and operation methods, as well as the electricity market's pricing and mechanism design. In this letter, we consider some of the implications of the 2023 RTAPS findings for the system planning and operation and electricity market.
\end{abstract}


\section{Introduction}
\IEEEPARstart{C}{laims} of a new superconducting material, LK-99 \cite{LK99-1,LK99-2}, have caught the attention of researchers in recent days. This superconductor not only works above room temperature but also works well under ambient pressure. The material is, therefore, known as the Room-Temperature Ambient-Pressure Superconductor (RTAPS). Leaving aside whether such materials have really achieved room temperature superconductivity, and regardless of the actual industrial application process of such materials, superconductivity seems to be no longer a dream but an important topic we need to start discussing today. 

Once proven effective, Room-Temperature Ambient-Pressure Superconductor utilization could lead to no energy loss in power transmission and distribution networks operated with Direct Current (DC) power, significantly improving electricity transmission efficiency, delivery reliability, grid flexibility, calculate efficiency, network intelligence, and renewable energy adoption. All foundations of modern power systems, including planning and operation, electricity market, power electronics, system security, and sustainability, will be profoundly affected. There are no more voltage regulation and stability issues in the power system when fully replaced with RTAPS. Existing systems and all traditional paradigms need to be rebuilt, and the existing research landscape should be reshaped. While enjoying superconductivity's great potential and vision, we will also face new challenges. For example, how to transfer to a power system based entirely on RTAPS in the case of existing traditional grid assets? How to deal with security issues and manage the new generation mix in the RTAPS-based power grid? 

Therefore, this paper explores the impact of superconductivity on power systems, including the opportunity and challenges. We discuss the RTAPS-based power flow method and security reconfiguration, the computational efficiency and cost implication, the pricing and settlement mechanisms, and the creation of new market transaction forms and products with the increasing competitiveness of renewable.   

\begin{figure}[t]
    \centering
    \includegraphics[width=\linewidth]{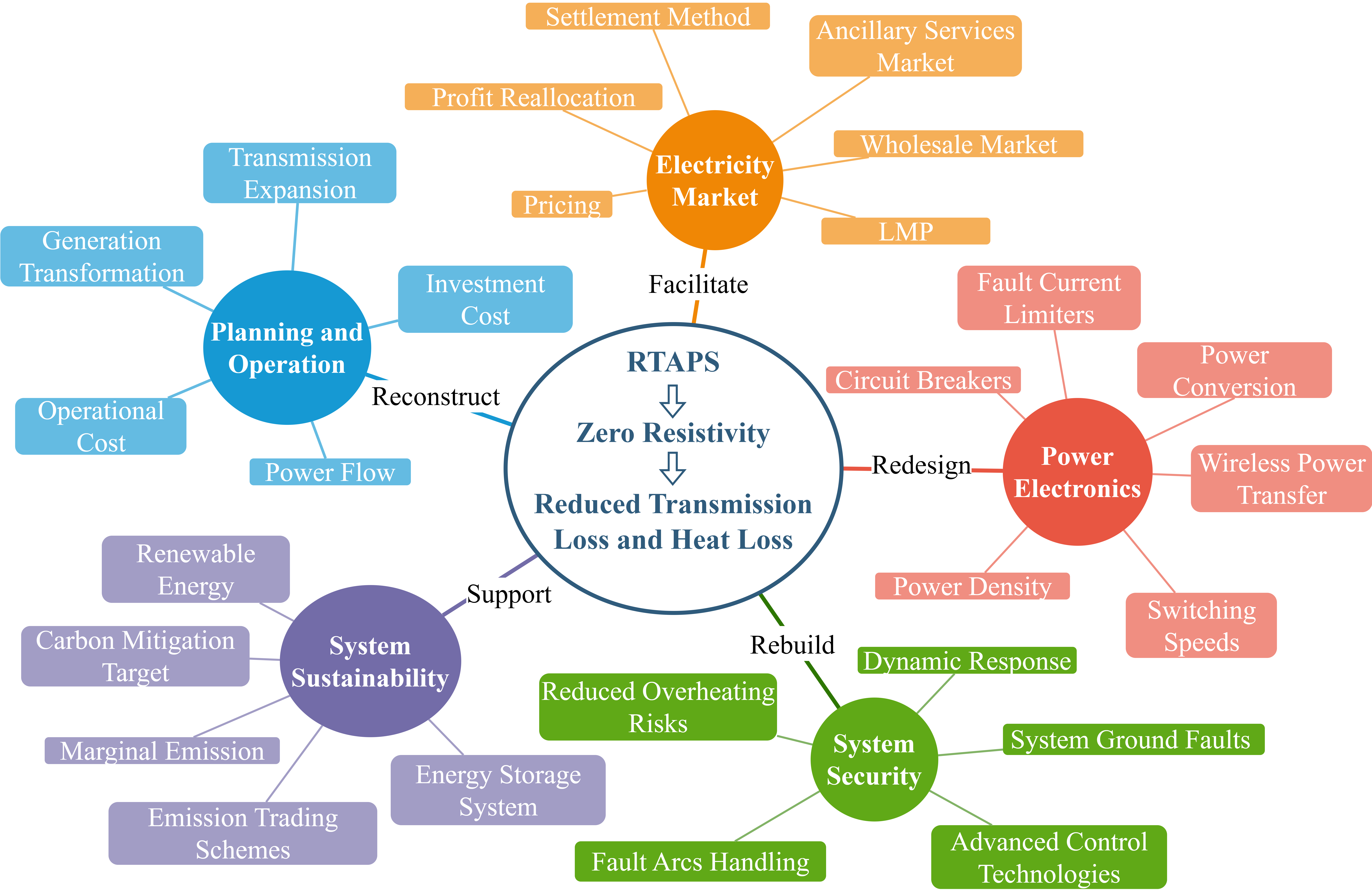}
    \caption{Overview of RTAPS-based power systems.}
    \label{fig:intro}
\end{figure}

\section{Background of Room-Temperature Ambient-Pressure Superconductor}
On July 22, 2023, Lee \textit{et al.} released two papers on their successful synthesis of a room temperature and atmospheric pressure superconductor (RTAPS) \cite{LK99-1, LK99-2}. The superconductor is named LK-99, which is a modified-lead apatite crystal structure, and its composition is (Pb$_{10-x}$Cu$_x$(PO$_4$)$_6$O ($0.9<x<1.1$)) \cite{LK99-2}. Diverging from previous related research, papers claim that LK-99 exhibits the Ohmic metal characteristic above the superconducting critical temperature \cite{LK99-2}. Furthermore, they provide the levitation phenomenon as the Meissner effect of a superconductor under conditions of room temperature (critical temperature is 126.85 degrees Celsius) and atmospheric pressure \cite{LK99-2}.
Zero resistivity, a characteristic of superconductors, may also be partly confirmed during the synthesis of the material.  
Since the release of these two striking articles, Griffin has provided some theoretical support for the material's superconducting properties \cite{Griffin}.
Lai \textit{et al.} show that the parent compound Pb$_{10}$(PO$_4$)$_6$O is an insulator, and the doping of copper induces an insulator-metal transition and thus a volume shrinkage \cite{Lai}.
Hou \textit{et al.} successfully observed zero resistance at a temperature of 110K, which may be very important evidence for the existence of superconductivity \cite{Hou}.
Given that circuit impedance plays a crucial role in traditional power systems, it's essential to consider the impact of the discovery of superconducting materials on it.

\section{Implications}
The application of superconducting materials in power grids has been discussed in academia for many years. However, due to technological limitations of the time, most applications were confined to smaller devices such as fault current limiters or ship propulsion motors \cite{3,4}. Malozemoff \cite{5} even expressed pessimistic views, suggesting that room-temperature superconductors are not suitable for power grids. In \cite{6}, the impact of High-Temperature Superconductivity (HTS) technology on power grids is extensively discussed. However, the article notes that the temperature of existing high-temperature superconductors is typically maintained between 100K and 140K under normal pressure conditions. Superconducting power devices often require supplemental cooling equipment such as liquid nitrogen, which increases cost and reduces convenience.

However, the RTAPS discovered this time does not have the above limitations. On the premise that the prospect of large-scale application is promising, this letter discusses the impact of RTAPS on power system issues from the aspects of system planning and operation, market, and emission.

\subsection{System planning and operation}

The application of RTAPS has a profound impact on system planning and operation. This includes changes to power flow calculation methods, computational efficiency, operation methods, and both long and short-term system costs.

\subsubsection{\textbf{Plow Flow Methods}}

The power flow calculation in traditional systems typically adopts AC or DC power flow. A typical AC power flow is illustrated below

\begin{align}
   {P_i} = {V_i}\sum\limits_{j = 1}^N {{V_j}\left( {{G_{ij}}\cos \left( {{\theta _i} - {\theta _j}} \right) + {B_{ij}}\sin \left( {{\theta _i} - {\theta _j}} \right)} \right)} \\
   {Q_i} = {V_i}\sum\limits_{j = 1}^N {{V_j}\left( {{G_{ij}}\sin \left( {{\theta _i} - {\theta _j}} \right) + {B_{ij}}\cos \left( {{\theta _i} - {\theta _j}} \right)} \right)} 
\end{align}

where $P_i$ and $Q_i$ are the active and reactive power of bus $i$; $V_i$ and $\theta _i$ denote the voltage magnitude and phase angle; $G_{ij}$ and $B_{ij}$ indicate the conductance and susceptance between bus $i$ and $j$. Given that superconductors still exhibit superconducting reactance characteristics in the AC environment, with the large-scale application of RTAPS to power networks, it's reasonable to assume that future power grids will be dominated by DC transmission. The system power flow model can be further simplified on the basis of DC power flow. 

\begin{align}
\sum S_{ij}+ P^{G} = D \\
\underline{S_{ij}}n^{TL}\leq S_{ij}\leq \overline{S_{ij}}n^{TL}
\end{align}

where $S_{ij}$ indicates the power flow on the transmission line; $P^G$ is generator power output; $D$ is the load; $n^{TL}$ denotes the candidate transmission line; $\overline{(\cdot)}$, $\underline{(\cdot)}$ are the upper and lower bound of variables. With a superconductor grid under the DC environment, the power flow still adheres to Kirchhoff's law, which states that the incoming power equals the outgoing power, as shown in (3). When considering the transmission expansion problem, it is only necessary to multiply the upper and lower limits by the number of candidate lines, as shown in (4). Despite the simpler model, it doesn't have calculation errors such as ignoring power loss and voltage drop. This can make the solution process faster and the calculation accuracy higher when solving large-scale problems. 

\subsubsection{\textbf{Operation methods}}

The application of superconducting materials could significantly alter the operation methods of existing systems, particularly the classification and definition of system security. Traditional networks, which are primarily based on AC transmission, face system dynamic issues such as voltage drop and frequency disturbance that affect system stability. In RTAPS-based systems, these stability problems may no longer exist. Besides stability, the traditional criteria for judging system security may no longer apply to RTAPS-based systems. In this new system environment, system security may depend more on issues such as the balance between source and load or whether changes in transmission line temperature could cause fluctuations in material resistance. Advanced protection and control technologies should be introduced to enhance the system's security as well.

\subsubsection{\textbf{Computational efficiency and grid intelligence}}

Given that changes in planning and operation methods will occur gradually as the system transforms, the initial impact of superconducting materials on the power system may be in terms of computational efficiency. The implications of superconductors on manufacturing, particularly the computer industry, will indirectly affect the power industry. Enhancements in computational efficiency and methodology could make real-time market clearing and system operation calculations possible. The solution time for nonlinear and mixed-integer problems in the current system will significantly decrease. The coupling of multi-energy systems, including electricity-gas, electricity-heat, and electricity-transportation, will extend beyond the academic realm into practical application. With the improvement of computing efficiency, the cost of training artificial intelligence systems is further reduced, especially for large-scale general artificial intelligence models and artificial intelligence agents. This is expected to directly improve the intelligence of the power system.

\subsubsection{\textbf{System Cost}}

Implementing RTAPS into power systems could significantly impact capital and operational costs. In the short term, system planning investment costs may substantially increase for retrofitting existing generators, transmission lines, and other devices. Second, regarding material costs, the price of commercial RTAPS products is still unclear. Based on the materials used in LK-99 as described in the currently published literature, the cost of equipment based on RTAPS may not necessarily increase significantly after large-scale application.

In the long run, the operating cost of the power system could be significantly reduced. The most significant cost savings would come from the drastically reduced energy losses in transmission and distribution. Secondly, RTAPS may have lower maintenance costs than traditional conductors since heat loss during power transmission will no longer exist, reducing the material's aging and the likelihood of device failure. However, the maintenance requirements of RTAPS are not yet known. This part of the cost may be increased.

Besides, RTAPS could also promote the cross-regional trading and transmission of renewable energy and energy storage systems and increase the activity and income of the electricity market. 

\subsection{Market}

Due to the reduction in network losses, the RTPAS will lead to significant changes in not only the pricing but also the market mechanism design of the modern electricity market.

\subsubsection{\textbf{Pricing and Profit Re-allocation}} RTAPS-based power systems will lead to modified Locational Marginal Pricing (LMP)\cite{8} or the new theory of spot pricing of electricity. Generally, LMP considers the transmission system’s constraints and the network losses, thereby determining the nodal electricity price in the power grid. However, transmission losses become negligible with RTAPS, and introducing these superconductors could also change transmission constraints and congestion pricing. It would impact LMP as the cost currently associated with these losses would no longer be a factor, potentially leading to more uniform pricing across a given network. The market dividends released by the disappearance of network losses should also be considered for the future profit re-allocation among the system operator, the market participants of the demand side, and the supply side.
\subsubsection{\textbf{Wholesale Market and Retail Market}} Regarding the wholesale market, market clearing pricing needs to be updated, including the calculation method for nodal marginal prices in the electricity spot market and the Time-of-use (TOU) pricing in the retail markets. Besides, superconductors could make transmitting power from renewable sources more feasible and will further motivate large-scale renewable energy sources' rapid integration into the power system. Integrating these distributed resources in market clearing raises numerous market design questions concerning the collaboration between Transmission System Operators (TSOs) and Distribution System Operators (DSOs). It may also lead to zero or even negative electricity prices, for which solutions must be proposed. It is necessary to rethink and redesign the bid-based market settlement approaches with a high penetration of variable renewable energy.

\subsubsection{\textbf{Ancillary Service Market}} In RTAPS-based power systems, the frequency and voltage regulation requests in the electricity ancillary services market may no longer exist since there is no network loss or line loss. Besides, the roles and investment plans of energy storage systems and traditional power generation units must be re-evaluated in the long run. Superconductors enable high-efficiency, large-scale, feasible, and widespread energy storage systems, making them ideal for grid stability applications. With the increased use of energy storage systems enabled by superconductors, demand may increase during off-peak hours, as these are when energy would be stored for later use. Also, RTAPS is pivotal in advancing nuclear fusion technology with higher efficiency and stable operations. While fusion offers a cleaner and safer form of nuclear energy, rigorous safety considerations in the power systems with nuclear energy are still essential to ensure its responsible development and implementation.

\subsubsection{\textbf{Innovative Market Services and Mechanisms}} With improved transmission capacity and reduced losses, superconductors could significantly improve grid flexibility, allowing electricity delivery over longer distances. This creates possibilities for long-distance cross-regional electricity trading and the application of physical power contracts. This may also enhance market profits and market liquidity. Various electricity and emission trading markets need to update their trading products, mechanisms, and settlement schemes to adapt to changes in transaction types and the increase in transaction scale. It's important to note that the full impact of room-temperature superconductors on market clearing and settlements would depend on many factors, including the pace of technology adoption and the cost and feasibility of implementing these materials on a large scale.

\subsection{Emission}
RTAPS may also have an impact on system emissions. For instance, applying RTAPS reduces transmission loss, improves power production efficiency, and reduces system emissions. It can also facilitate the large-scale grid integration of renewable energy and expedite the decarbonization transformation of the system. Superconductors could potentially make renewable energy sources more competitive, increasing the market share of renewable energy providers. Furthermore, the existing demand-side carbon emission calculation model may need to be adjusted to accommodate the characteristics of the RTAPS-based power systems.

\subsection{Recommendations}

The introduction of RTAPS will prompt power grid operators to reevaluate and reoptimize the structure and mechanism of the power system and market. The optimized grid and market will become more flexible and efficient, adapting to future energy transitions and power supply and demand changes. Despite the numerous advantages of RTAPS, there are still technological and economic challenges in their commercial application. Issues such as manufacturing costs, reliability, long-term stability, and limited current-carrying capacity of RTAPS still need to be addressed, and corresponding production and supply chain systems need to be established as well \cite{7}. 

Since most countries have completed their construction of basic power systems and markets, blind reconstruction may cause a lot of waste. Therefore, a mixed application strategy of traditions and RTAPS can be adopted. As the power losses in traditional systems are primarily concentrated in high-voltage long-distance transmission and low-voltage customer-side distribution networks, these two levels will be the first retrofit networks following the large-scale industrial application of RTAPS. The frameworks and mechanisms of RTAPS-based or traditional systems can be flexibly selected according to the needs of different regions. As RTAPS technology develops and decreases costs, its applicability in power systems will gradually expand.

\section{Conclusion}

In 2023, LK-99 could become the first RTAPS with a disclosed and reproducible preparation method. The development of superconductors will profoundly impact system planning and operation, markets, and pricing mechanisms in the near future. Specifically, this letter posits that power flow models, planning and operation costs, system security reconfiguration, price settlement, profit allocation, and market mechanisms will need to be adjusted accordingly to accommodate the forthcoming era of RTAPS-based power systems.

% Biography text here without a photo.

% \begin{IEEEbiography}[{\includegraphics[width=1in,height=1.25in,clip,keepaspectratio]{fig1.png}}]{IEEE Publications Technology Team}
% In this paragraph you can place your educational, professional background and research and other interests.\end{IEEEbiography}

\end{document}